\documentstyle{article}
\begin{document}

\author{{\bf Edgardo T. Garcia Alvarez and Fabi\'an H. Gaioli} \and {\it %
Instituto de Astronom\'\i a y F\'\i sica del Espacio, } \\
{\it C.C. 67, Suc. 28, 1428 Buenos Aires, Argentina and}\\
{\it Departamento de F\'\i sica, Facultad de Ciencias Exactas y Naturales,%
}\\
{\it Universidad de Buenos Aires, 1428 Buenos Aires, Argentina }}
\title{Feynman's proper time approach to QED}
\maketitle

\begin{abstract}
The genesis of Feynman's original approach to QED is reviewed. The main
ideas of his original presentation at the Pocono Conference are discussed
and compared with the ones involved in his action-at-distance formulation of
classical electrodynamics. The role of the de Sitter group in Feynman's
visualization of space-time processes is pointed out.
\end{abstract}

\textwidth=18cm \textheight=23.2cm \oddsidemargin=0.6cm \headheight=-2.0cm


\smallskip
{\it ``I had used it to formulate quantum electrodynamics. I invented it
to do that. It was in fact the mathematical formulation that I expressed at
the Pocono Conference --that was this crazy language. Dates don't mean
anything. It was published in 1951, but it had all been invented by 1948. I
called it the operator calculus.''} R.P. Feynman \cite{M}

\medskip\ 

\section{Introduction}

We devote this article to review Feynman's ideas since we think that there
is an incomplete knowledge and appreciation of them in quantum field theory
(QFT). The reason is that most people have only payed attention to Feynman's
papers of 1949 \cite{F49a,F49b}, and learned the derivation of Feynman's
rules for QED from Dyson's paper of 1948 \cite{Dy}. So some important ideas
involved in the papers of 1950 \cite{F50} and 1951 \cite{F51} --in which
Feynman exposed his own way-- were practically forgotten. Fortunately some
recent works of Schweber \cite{S86,S94} and Mehra \cite{M} have rescued
them. Here we present a brief account of these ideas to prepare the
discussion about a hidden de Sitter invariance in QED which can be taken as
the starting point to reformulate it \cite{G96,G97}.

\section{The Wheeler-Feynman theory}

The genesis of Feynman's approach to QED has its origin in his celebrated
paper with Wheeler \cite{W}. There they considered an action-at-distance
theory of Fokker's type. Their purpose was to formulate a classical theory
free from the well-known divergences due to self-interaction, so they
originally assumed that the charges do not act on themselves. The idea of an
action-at-distance theory was pursued again by Feynman in his article ``A
Relativistic Cut-off for Classical Electrodynamics'' \cite{F48} but this
time self-interaction was allowed. He considered an action of the form

\begin{equation}
S=\sum_nm_n\int ds_n+\frac 12\sum_{nm}e_ne_m\int \int f(s_{nm}^2)dx_\mu
^ndx^{m\mu },  \label{F48}
\end{equation}
where $s_{nm}^2=(x_\mu ^n-x_\mu ^m)(x^{n\mu }-x^{m\mu }),$ and $x_\mu ^n,$ $%
s_n$, $m_n,$ and $e_n$ are the coordinates, proper time, mass, and charge of
the $n$-particle respectively. $f$ is an invariant function of $s_{nm}^2,$
which behaves like the Dirac delta, $\delta ,$ for large distances but has a
cut-off for small ones. This was the trick he used in this case to avoid the
divergences of the self-interaction. The variation of the action (\ref{F48})
leads to the Lorentz force-law for the particles 
\begin{equation}
m_n\frac{d^2x_\mu ^n}{ds_n^2}=e_n\frac{dx_\mu ^n}{ds_n}\sum_mF_m^{\mu \nu
}(x_n),  \label{LF}
\end{equation}
where $F_m^{\mu \nu }(x)=\partial ^\mu A_m^\nu -\partial ^\nu A_m^\mu ,$ is
the field caused by the $m$-particle corresponding to the potential

\begin{equation}
A_m^\nu (x_n)=e_m\int f(s_{nm}^2)dx_\mu ^m.  \label{A}
\end{equation}
As $\delta (s_{nm}^2)$ is a Green function of d'Alembertian operator with
the boundary condition corresponding to the half-advanced plus half-retarded
solution, $A_m^\nu $ satisfies Maxwell equations at long distances of the
sources.

Concluding his 1948 article \cite{F48} Feynman described another important
result. He noticed that the explicit covariant formulation of classical
electrodynamics allowed him to introduce the notion of antiparticles. In
fact form Eq. (\ref{LF}) it follows that the proper time reversal is
equivalent to conjugate the sign of the charge. Following this idea,
suggested to him by Wheeler in 1941, Feynman described pair production in
external fields.

\section{ The Pocono Conference}

At the Pocono Conference (1948) Feynman gave a dissertation about an
``Alternative Formulation of Quantum Electrodynamics.'' The notes of
Feynman's talk taken by Wheeler were recently published by Schweber \cite
{S86,S94}. These important works of Schweber clarify the genesis of
Feynman's ideas, which is not so explicit in the most widely known published
material (the two papers of 1949 \cite{F49a,F49b} and the Lectures on QED
given by Feynman at Caltech in 1953 \cite{F61}), and only appeared in the
appendices of his later publications \cite{F50,F51}.

In his dissertation Feynman introduced the following parametrization of the
Dirac equation in an external electromagnetic field 
\begin{equation}
-i\frac{\partial \Psi (x,s)}{\partial s}=\gamma ^\mu (i\partial _\mu -eA_\mu
)\Psi (x,s),  \label{FeyA}
\end{equation}
where $A_\mu $ is the electromagnetic potential. However it is important to
remark that he introduced the fifth parameter in a purely formal way and did
not explicitly attributed any physical meaning to it. He denoted it as $w$,
but we prefer to call it $s$ because, in the classical limit, it can be
identified with the proper time \cite{ap95b}.

Equation (\ref{FeyA}) is a Schroedinger equation in the invariant parameter $%
s$ which labels the evolution of states out of the mass-shell. The
mass-shell condition is satisfied by stationary states of mass $m,$ $\Psi
(x^\mu ,s)=\psi _m(x^\mu )e^{ims},$ where $\psi _m(x^\mu )$ are solutions of
the Dirac equation for a system with mass $m,$

\begin{equation}
\left[ \gamma ^\mu (i\partial _\mu -eA_\mu )-m\right] \psi _m(x)=0.
\label{Dir}
\end{equation}

The key idea of Feynman \cite{F51} was that by Fourier transforming in $s$
any solution $\Psi (x,s)$ of Eq. (\ref{FeyA}) a solution $\psi _m(x)$ of Eq.
(\ref{Dir}) can be obtained$,$ namely

\begin{equation}
\psi _m(x)=\int_{-\infty }^{+\infty }\Psi (x,s)e^{-ims}ds.  \label{psi}
\end{equation}
Hence the Fourier transform of the retarded Green function $G(x,x^{\prime
},s)$ of Eq. (\ref{FeyA}) $(\pi _\mu =i\partial _\mu -eA_\mu )$,

\begin{equation}
\left( \gamma ^\mu \pi _\mu -i\frac \partial {\partial s}\right)
G(x,x^{\prime },s)=\delta (x,x^{\prime })\delta (s),  \label{green}
\end{equation}
enables one to derive the corresponding mass-shell Green function $%
G_m(x,x^{\prime }),$ {\it i.e.}

\begin{equation}
\left( \gamma ^\mu \pi _\mu -m\right) G_m(x,x^{\prime })=\delta (x,x^{\prime
}).  \label{shell}
\end{equation}
That is $\left[ {\rm as }G(x,x^{\prime },s)=0, {\rm for }s\leq 0\right] $

\begin{equation}
G_m(x,x^{\prime })=\int_0^{+\infty }G(x,x^{\prime },s)e^{-ims}ds.
\label{GsGm}
\end{equation}
Taking into account that the off-shell retarded Green function is $%
G(x,x^{\prime },s)=-i\theta (s)\left\langle x\left| e^{i\gamma ^\mu \pi _\mu
s}\right| x^{\prime }\right\rangle $ and using the formal identity $%
i/(a+i\epsilon )=\int_0^\infty \exp [is(a+i\epsilon )]ds$ for $a=\gamma ^\mu
\pi _\mu -m$, one immediately sees that such retarded boundary condition for 
$G(x,x^{\prime },s)$ naturally leads to the Feynman $i\epsilon $
prescription for avoiding the poles in the on-shell Green function:

\begin{equation}
G_m(x,x^{\prime })=\left\langle x\left| \frac 1{\gamma ^\mu \pi _\mu
-m+i\epsilon }\right| x^{\prime }\right\rangle .  \label{Fbc}
\end{equation}
This formal trick allowed Feynman to discuss external field problems of QED
keeping it up at a first-quantized level.

In order to discuss the radiative processes Feynman wrote down a closed
expression for the off-shell amplitude of a system of spin-half charges

\begin{eqnarray}
&&\Psi (x_1,...x_n,s_1,...s_n)  \nonumber  \label{pocono} \\
&=&\exp \left\{ -i\left[ \sum_n\int_0^s\gamma _n^\mu (s_n^{\prime })\pi _\mu
^n(s_n^{\prime })ds_n^{\prime }\right. \right.  \nonumber \\
&&\left. \left. +\sum_{nm}\frac{e_ne_m}2\int_0^{s_n}\int_0^{s_m}\gamma
_n^\mu (s_n^{\prime })\gamma _\mu ^m(s_m^{\prime })\delta _{+}\{[x_\mu
^m(s_m^{\prime })-x_\mu ^n(s_n^{\prime })]^2\}ds_n^{\prime }ds_m^{\prime
}\right] \right\}  \nonumber  \label{pocono} \\
&&\times \Psi (x_1,...x_n,0,...0).  \label{pocono}
\end{eqnarray}

Equation (\ref{pocono}), after an integration in the parameters $s_n,$ as in
(\ref{psi}), allowed him to describe any process of QED in which only
virtual photons are present. From that one he deduced the corresponding
diagrammatic expansion of QED. He explicitly showed how to derive the vertex
diagram to order $e^2$ and lowest order in the external field and the
self-energy terms to order $e^2.$

Feynman proposed Eq. (\ref{pocono}) in 1948 at the Pocono Conference, but he
did not have a formal derivation of it, whose justification was given later
in his paper of 1951.\footnote{%
See the comments in footnote 19 of Feynman's 1951 paper \cite{F51}.} He was
led for the intuition gained in his previous works in which {\it via} the
path integral method he computed the transition amplitude between processes
in which the initial and final states correspond to the radiation
oscillators in the vacuum. In such a case he could eliminate the radiation
oscillators in the action of scalar electrodynamics and obtain and effective
action similar to (\ref{F48}) but with the Green function $\delta
_{+}(s_{nm}^2)$ instead of $f(s_{nm}^2).$ At that time the path integration
for spinning particles was unknown. So to deal with the spin-half case
Feynman had to invent an alternative formalism, his operator calculus \cite
{F51}. The idea is that the order in which the operators act is determined
by the order of some associated parameters, which in the case of expression (%
\ref{pocono}) are the evolution parameters $s_{n.}$ The exponential in Eq. (%
\ref{pocono}) is the evolution operator for a system of interacting charges.
In the first sum we recognize the evolution operator of the external field
problem for each charge. The second bilinear expression resembles the one
which appears in the action-at-distance theory. When writing Eq. (\ref
{pocono}) Feynman just associated the proper time velocities $\frac{dx^\mu }{%
ds}$ of his effective action for the scalar case with the Dirac matrices $%
\gamma ^\mu .$

As was emphasized by Schweber \cite{S86} Feynman was harshly criticized by
the audience. Bohr misunderstood his pictorial use of particle trajectories
prohibited by the uncertainty principle. Dirac did not receive a
satisfactory answer to his question about the unitarity of the
transformation (\ref{pocono}). This was the reason why he felt somewhat
discouraged and decided to publish his result in the way he did in his two
celebrated papers of 1949.

\section{The papers of 1949}

In 1949 Feynman \cite{F49a,F49b} published two papers just presenting the
rules of QED by means of some plausible intuitive arguments. He avoided the
use of path integration or his operator calculus, both techniques weird at
that time.\footnote{%
In the papers of 1949 Feynman avoided the use of path integrals for
evaluating the Green functions, and only used the Stueckelberg\cite{S41}
interpretation for antiparticles on the mass shell to determine their
boundary conditions.} It is clear that he decided to conceal his thoughts in
order to be better understood. He didn't reveal them until he wrote his
papers of 1950 and 1951 \cite{F50,F51}.

Some time before the papers of 1949 appeared in the Physical Review, Dyson
published a paper \cite{Dy} about the equivalence of Feynman's rules with
the Tomonaga and Schwinger formulation. Dyson's paper had a double effect.
On the one hand it legitimized the use of Feynman diagrams in the community,
which quickly became more and more popular as an easier calculation device.
But on the other hand, it made people forget the ideas presented by Feynman
at Pocono in 1948, which was the original way in which Feynman had derived
the rules of QED. These ideas are very important because they represent an
alternative formulation of QED to the one given by Tomonaga, Schwinger, and
Dyson in the framework of QFT. Moreover, Dyson's paper does not imply that
both theories are strictly equivalent. It only implies that both theories
have in common the same diagrammatic expansion of the $S$ matrix on the mass
shell.

\section{The de Sitter invariance of QED}

The purpose of this section is to show that the Pocono formulation is richer
than QFT.

In effect by multiplying on the left by $\gamma ^5$ we can rewrite Eq. (\ref
{FeyA}) as a five-dimensional massless Dirac equation ($A=0,1,2,3,5$)
\begin{equation}
\Gamma ^A(i\partial _A-eA_A)\Psi =M\Psi ,\hspace{0.3in} M=0,
\label{irrep}
\end{equation}
where $A_A=(A_\mu ,$ $A_5=0)$ and $\Gamma ^\mu =\gamma ^5\gamma ^\mu $ and $%
\Gamma ^5=\gamma ^5=\gamma ^0\gamma ^1\gamma ^2\gamma ^3$ satisfy the Dirac
algebra

\begin{equation}
\Gamma ^A\Gamma ^B+\Gamma ^B\Gamma ^A=2g^{AB},
\end{equation}
corresponding to a five-dimensional manifold of coordinates $x^A=(x^\mu ,x^5)
$ endowed with a super-Minkowskian metric $g^{AB}={\rm diag}(+,-,-,-,-).$
We have identified $x^5$ with $s,$ due to as $M=0,$ the five-line element 
\begin{equation}
dS^2=g^{AB}dx_Adx_B=g^{\mu \nu }dx_\mu dx_\nu -(dx^5)^2  \label{ds}
\end{equation}
vanishes. In other words, the super-particles (off-shell particles) go at
the speed of light ($dS=0$)

\begin{equation}
\frac{dx^\mu }{dx^5}\frac{dx_\mu }{dx^5}=1,
\end{equation}
in analogous sense that the standard particles go at the speed of light $%
(ds=0)$ $\left( \frac{d{\bf x}}{dt}\right) ^2=1,$ when $m=0.$\footnote{%
Note however that $x^5$ is arbitrary in principle, and the identification of 
$x^5$ with $s$ is only valid in the equations of motions whose solutions
lies on the super-light cone.}

The group of linear transformation of coordinates $x^{A^{\prime }}=L_{\  
B}^Ax^B+C^A$ which leave $dS^2$ invariant is the de Sitter group. We have
rewritten (\ref{FeyA}) in the form (\ref{irrep}) to explicitly show that it
is de Sitter covariant. In Sec. 3 we have outlined the way in which he could
develop the formalism of QED starting from (\ref{FeyA}). So here we
concentrate our attention on some particular points related to the
interpretation of the formalism.\footnote{%
The connection of this formalism with classical theories of spinning
particles is discussed in Ref. \cite{spin}.}

The Noether current associated with Eq. (\ref{FeyA}) determines the
indefinite Hermitian form

\begin{equation}
\left\langle \Psi |\Phi \right\rangle =\int d^4x\overline{\Psi }(x)\Phi (x).
\label{pe}
\end{equation}
The covariant Hamiltonian or mass operator ${\cal H=}$ $\gamma ^\mu \pi
_\mu $ is ``self-adjoint'' in (\ref{pe}) so the evolution operator $e^{i%
{\cal H}s}$ is ``unitary.'' This answers Dirac's question we quoted in
Sec. 3.

The evolution of any operator $A$ in the Heisenberg picture is given by

\begin{equation}
\frac{dA}{ds}=-i[{\cal H},A],  \label{dbe}
\end{equation}
which is the proper time derivative originally proposed by Beck \cite{beck}
in 1942.\footnote{%
During the last fifty years the Feynman parametrization and the Beck proper
time derivative were rediscovered or discussed by many authors (for a list
of references see Refs. \cite{ap95a,G98}).} In this formalism, the
coordinate time $x^0$ has been elevated to the status of an operator
canonically conjugated to the energy $p^0$. Their commutation relation and
the standard canonical commutation relation for the three-position and
-momentum can be summarized in the covariant commutation relation

\begin{equation}
\lbrack x^\mu ,p^\nu ]=-i\eta ^{\mu \nu }.  \label{xpop}
\end{equation}
Using (\ref{dbe}) for $A=x^\mu $ and (\ref{xpop}) we obtain the covariant
generalization of Breit's formula

\begin{equation}
\frac{dx^\mu }{ds}=\gamma ^\mu ,  \label{Breit}
\end{equation}
which justifies the association made by Feynman in writing Eq. (\ref{pocono}%
).

It can be proved that at a semiclassical level \cite{ga96a}

\begin{equation}
{\rm sign}\left[ \overline{\Psi }(x,s)\Psi (x,s)\right] ={\rm sign}%
\frac{dx^0}{ds}.
\end{equation}
It means that, according to the Stueckelberg \cite{S41} interpretation,
super-particles and super-antiparticles states have positive and negative
``norms'' respectively. This is the root of the indefinite character of the
``inner product.'' Frequently, this fact is considered as an anomaly of the
theory, but it is a consequence of having a canonical description with
particles and antiparticles on the same footing \cite{G98}.

We have appealed to the Stueckelberg interpretation for antiparticles at the
classical level. Let us see that it naturally arises in the formalism
according to the more familiar notion based on charge conjugation. The
operation that conjugates the charge in Eq. (\ref{FeyA}) is \cite
{Han91,Han94,G96,G97}

\begin{equation}
C\Psi (x,s)=c\Psi (x,-s),
\end{equation}
where $c=\gamma ^5K$ is the standard charge conjugation operator. The
remarkable point is that this operation coincides with the $s$-time reversal
operation in the Wigner sense

\begin{equation}
C=S.  \label{wig}
\end{equation}
The identity (\ref{wig}) is the quantum analogue of the Feynman observation
quoted in Sec. 2. That is, charge conjugation in the Lorentz force law is
equivalent to an inversion of the sign of $\frac{dx^0}{ds}.$

The comparison between Pocono formula (\ref{pocono}) and the classical
Fokker's action (\ref{F48}) clearly shows\footnote{%
Note that we can rewrite in action (I-1) $dx_\mu ^n=\frac{dx_\mu ^n}{ds_n}%
ds_n$ and identify $\frac{dx_\mu ^n}{ds_n}$ with $\gamma _\mu $ and $m_n$
with the mass operator $\gamma _\mu ^n\pi ^{n\mu }.$} that the classical
limit of the Pocono theory is an action-at-distance formulation of
electrodynamics with the Feynman boundary conditions for the on-shell Green
function of the d' Alembertian

\begin{equation}
\partial ^\mu \partial _\mu \delta _{+}\left[ (x_\mu -x_\mu ^{\prime
})^2\right] =\delta (x,x^{\prime }).
\end{equation}
Let us show that such boundary conditions correspond to the retarded ones in
a five-dimensional electromagnetic theory\footnote{%
Five-dimensional electromagnetism is also discussed by Saad, Horwitz, and
Arshansky \cite{Saad} and Shnerb and Horwitz \cite{Shnerb}.}

\begin{equation}
\partial ^A\partial _AA^B=j^B,\hspace{0.3in} \partial _AA^A=0.
\end{equation}
The off-shell Green function $D(x,x^{\prime },s)$ of the wave operator $%
\partial ^A\partial _A,$

\begin{equation}
\left[ \partial ^\mu \partial _\mu -\frac{\partial ^2}{\partial s^2}\right]
D(x,x^{\prime },s)=\delta (x,x^{\prime })\delta (s),
\end{equation}
is related to the spin 1/2 Green function by

\begin{equation}
G(x,x^{\prime },s)=-\left[ \gamma ^\mu i\partial _\mu +i\frac \partial
{\partial s}\right] D(x,x^{\prime },s).  \label{GD}
\end{equation}
Integrating Eq. (\ref{GD}) between $s=0$ and $s=\infty $ and assuming the
retarded boundary conditions $\left[ D(x,x^{\prime },\infty )=D(x,x^{\prime
},0)=0\right] $ we have

\begin{equation}
G_{m=0}(x,x^{\prime })=-\gamma ^\mu i\partial _\mu D_{m=0}(x,x^{\prime }),
\label{relacion1}
\end{equation}
where

\begin{equation}
D_{m=0}(x,x^{\prime })=\int_0^{+\infty }D(x,x^{\prime },s)ds.  \label{DsDm}
\end{equation}
As we know from (\ref{Fbc}) that $G_{m=0}(x,x^{\prime })$ satisfies
Feynman's boundary conditions, $D_{m=0}(x,x^{\prime })$ also satisfies it.
But multiplying (\ref{relacion1}) by $\gamma ^\mu i\partial _\mu $ and using
(\ref{shell}) we see that $D_{m=0}(x,x^{\prime })=\delta _{+}\left[ (x_\mu
-x_\mu ^{\prime })^2\right] .$

Notice also that (\ref{DsDm}) is the analogue of (\ref{GsGm}). The
exponential factor does not appear in this case because the photon mass is
zero.

\section{Final remarks}

Once a line of thought is installed in any field of science it has a very
high possibility of success --greater than any other better alternative
point of view-- due to the great number of thinkers playing with the leading
ideas. In the case of QED, most physicists have learnt Feynman's rules from
Dyson works, and probably they did not know very much about Feynman's
original ideas.\footnote{%
A\ notable exception is Nambu  \cite{nambu}, who based on Feynman's 1949
papers and an earlier work of Fock  \cite{fock} also realized of the
importance of proper time in QED.} Of course, in part this because Feynman
himself concealed part of his thoughts in his papers of 1949. It is clear
that Feynman did not find a framework to justify all his intuitions and
finally he plunged into an orthodox point of view (albeit not completely).
Remind that when Dyson told him about his paper \cite{Dy}, Feynman had no
qualms about giving him a free hand to publish his ideas before Feynman had
published them himself. Feynman just said: ``Well, that's great! Finally I
am respectable.''\footnote{%
See Mehra's book \cite{M}, Chap. 13.}

In this work we have presented an intent to rescue the original ideas since
from Feynman's view we can go beyond QFT. As we have shown, while QFT rests
on Poincar\'e invariance the Pocono formulation hides the de Sitter
invariance of QED.

\bigskip\ \ 

\noindent {\bf Acknowledgments}

\smallskip\ 

We are grateful to Bill Schieve for his kindly invitation to the University
of Texas and his encouragement for writing this paper and to Larry Horwitz
for reading the manuscript. We also want to express a special acknowledgment
to Leonid Burakovsky and Terry Goldman for his warm hospitality at Los
Alamos.

\end{document}